# Dynamical fermion simulations using Lüscher's local bosonic action algorithm

Mike Peardon [a] (UKQCD collaboration)

[a]Department of Physics and Astronomy, University of Edinburgh, Edinburgh, EH9 3JZ, Scotland, UK.

The two-flavour Schwinger model is used to test the local boson action algorithm of M Lüscher. The autocorrelation time is found to rise linearly with the number of auxiliary boson fields. An extension to the algorithm is proposed which exactly reproduces the full dynamical partition function for any number of boson fields.

## 1. Introduction

### 1.1. The Schwinger model

The Schwinger model (2d QED) with staggered fermions provides a computationally simple testing ground for dynamical fermion algorithms. The theory is asymptotically free, so results relating to algorithm performance should be similar to simulations of QCD.

The model provides a useful probe of the low-eigenvalue behaviour of the algorithm which is particularly critical for the local boson method since the polynomial approximation breaks down here. The presence of approximate zero modes [3] (AZMs) associated with the topologically charged sectors of the theory prove difficult to handle.

### 1.2. Lüscher's local bosonic algorithm

The algorithm [1] exploits a bosonic path integral to simulate the fermion determinant of the full theory via a polynomial approximation to the inverse of the fermion matrix. The partition function for the Schwinger model with 2 flavours of fermions is

$$Z = \int \mathcal{D}U \, \det Q \, e^{-S_G[U]} \quad (1)$$

$Q$ is an hermitian version of the fermion matrix, scaled so it has eigenvalues in the range $(-1, 1)$

$$Q = \frac{\gamma_5(\slashed{D} + m)}{\lambda_{\max}} \quad (2)$$

Consider an $n^{\text{th}}$ order polynomial approximation, $\mathcal{P}_n(s)$ to $1/s$ where $0 < s < 1$. The polynomial has $n$ roots in the complex plane, and can be written

$$\mathcal{P}_n(s) \propto \prod_{i=1}^{n} (s - z_i) \quad (3)$$

The optimal choice of roots comes from an analysis of Chebyshev polynomials. The lower scale for accelerated convergence is set by the choice of parameter, $\epsilon$.

With a set of roots, $\{z_i\}$ the determinant of $Q$ can be replaced by

$$\det Q^2 \propto \frac{1}{\det \mathcal{P}_n(Q^2)} = \frac{1}{\prod_{i=1}^{n} \det(Q^2 - z_i)} \quad (4)$$

For the two flavour theory, $\sqrt{\det \mathcal{P}_n(Q^2)}$ is required. The hermitian fermion matrix of (2) has eigenvalues in $\pm$ pairs which leads to a degeneracy in pairs of determinants in the product of (4). The square root is taken by removing one of the determinants in each pair.

A set of auxiliary bosons is introduced with partition function, $Z_{LA}$ which mimics the fermion determinant

$$Z_{LA}[U] = \int \mathcal{D}\chi_f \mathcal{D}\chi_f^* \, e^{-\chi_f^*(Q-\sqrt{z_i^*})(Q-\sqrt{z_i})\chi_f} \quad (5)$$

so that $\det Q \approx Z_{LA}$

Note that the number of auxiliary fields is half the order of the polynomial for two flavour simulations.

### 1.3. Implementation

Updates are as follows:

- the boson fields are updated from a (gaussian) heatbath, since they have a local quadratic action.

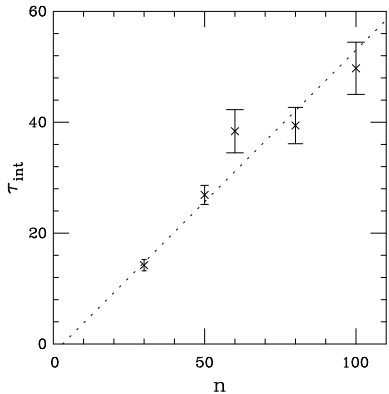

Figure 1. The plaquette integrated autocorrelation time vs. order of polynomial ($\epsilon = 0.005$)

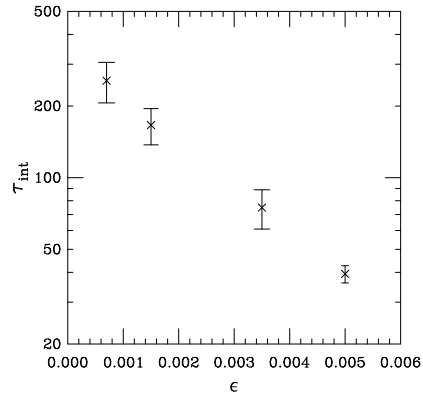

Figure 2. The plaquette integrated autocorrelation time vs. $\epsilon$ ($n = 80$)

- gauge field updates are over-relaxation only. Ergodicity is ensured as the total action of the theory is updated stochastically when the boson fields are updated. There is little advantage in stochastic updates of the gauge fields.

## 2. Autocorrelation times

The integrated autocorrelation time for the plaquette is shown as a function of $n$ and $\epsilon$ for a $16 \times 16$ lattice, bare fermion mass = 0, $\beta = 3.0$.

From figure 1 we see that $\tau_{\text{int}} \propto n$. The CPU autocorrelation time (assuming the boson updates dominate) will therefore rise in proportion to $n^2$. An estimate for the number of fields required to simulate fermions of mass $m$, is $n \propto 1/m$. This implies the cost of the algorithm will rise at least as $1/m^2$.

Figure 2 shows that the autocorrelation time also rises as the lower scale of the approximation is reduced. The polynomial roots lie on an ellipse with focii at $\epsilon$ and 1 and with minor axis length $2\sqrt{\epsilon}$. As $\epsilon$ is reduced, the smallest eigenvalues of each boson coupling matrix will decrease like $\epsilon$ so the autocorrelation time will rise approximately like $1/\epsilon$

The increase is not alleviated by increasing the denominator of (2). For staggered fermions, (unlike Wilson fermions [2]) the spectrum contains few eigenvalues near to $s = 1$.

## 3. LARD - Local Action + Reduced Determinant

The full partition function of the theory is

$$Z = \int \mathcal{D}U \; \det Q \, e^{-S_G[U]}$$

$$= \int \mathcal{D}U \; \frac{\det Q \sqrt{\det \mathcal{P}(Q^2)}}{\sqrt{\det \mathcal{P}(Q^2)}} \, e^{-S_G[U]} \quad (6)$$

Employing the partition function of the local bosonic theory, (5),

$$Z = \int \mathcal{D}U \; \det Q \; \sqrt{\det \mathcal{P}(Q^2)} \; Z_{LA}[U] \, e^{-S_G[U]} \quad (7)$$

If the polynomial approximation were exact, the product of the two determinants in (7) would be 1. In practise, use of a computationally accessible number of polynomial terms ($\equiv$ number of boson fields) means there is some error in the polynomial and hence fluctuation in the product over configurations. The product, $\mathcal{O}_{\text{RD}}$ quantifies the error in the polynomial on every configuration. $\mathcal{P}$ is a polynomial, hence it is simultaneously diagonisable with $Q$ so

$$\mathcal{O}_{\text{RD}} = \sqrt{\det Q^2 \mathcal{P}(Q^2)} = \prod_i^{V/2} \lambda_i \mathcal{P}(\lambda_i) \quad (8)$$

where $\lambda_i$ are one of each of the degenerate pairs of eigenvalues of $Q^2$ The reduced determinant, $\mathcal{O}_{\text{RD}}$, has the following properties:

1. the local boson method is "quenching" this operator

2. as $\mathcal{P} \to 1/s$, $\mathcal{O}_{\rm RD} \to 1$

3. significantly reduced fluctuations cf. $\det Q$

Writing the partition function as,

$$Z = \int \mathcal{D}U\, \mathcal{O}_{\rm RD}[U]\, Z_{LA}[U]\, e^{-S_G[U]} \quad (9)$$

the algorithm will exactly reproduce the full unquenched partition function, if a Metropolis accept/reject step is included to incorporate the operator, $\mathcal{O}_{\rm RD}$

### 3.1. Implementation

The updates of the gauge and boson fields are identical to those used in the original algorithm, with the additional constraint that the forward and backward rates for updates are identical. This is required to ensure detailed balance for the entire Markov step.

The exact algorithm is

- calculate $\mathcal{O}_{\rm RD}$ on the gauge configuration $\{U\}$

- update the bosons and gauge fields according to the local boson method

- recalculate $\mathcal{O}_{\rm RD}$, and accept the new gauge configuration, $\{U'\}$, with probability

$$P_{\rm acc} = \min[1, \frac{\mathcal{O}_{\rm RD}[U']}{\mathcal{O}_{\rm RD}[U]}]$$

To calculate $\mathcal{O}_{\rm RD}$ fully, the fermion matrix must be diagonalised. This is a computationally intensive step. In practise, for a suitable choice of the parameter, $\epsilon$ the dominant contribution to fluctuations in $\mathcal{O}_{\rm RD}$ come from the lowest lying modes of the fermion matrix. As a result, the eigenvalues product of (8) can be cut-off at some lower bound.

If all eigenvalues are extracted or a stochastic estimate of the determinant is used, the fermion mass appearing in the auxiliary boson coupling matrix need not be equal to the true fermion mass. The mass used in the local boson partition function can be tuned to maximise the acceptance probability of the Metropolis test.

### 3.2. Acceptance rates

The acceptance probability of the global Metropolis step is high enough to allow the use of low $n$ polynomials. Table 1 shows how the acceptance probability alters as the number of sweeps over the lattice between Metropolis tests is increased.

Table 1
Acceptance probabilities for global Metropolis step $16 \times 16$ lattice, $\beta = 3.0$ ($n = 16, \epsilon = 0.05$)

| # Sweeps per acc/rej | Acceptance rate |
| --- | --- |
| 1 | $0.50 \pm 0.01$ |
| 2 | $0.42 \pm 0.01$ |
| 4 | $0.30 \pm 0.01$ |
| 8 | $0.24 \pm 0.01$ |

### 4. Discussion

The addition of the accept/reject step provides a possible method of reducing the constraint of the linear rise in autocorrelation times with the number of bosons. It also guarantees exactness for any choice of polynomial. This is of importance in Schwinger model simulations due to the appearance of the topological AZMs.

Whether the method improves performance for $4d$ theories is less certain, as the cost of calculating $\mathcal{O}_{\rm RD}$ (either exactly or stochastically) may be prohibitively expensive.

Comparisons of performance against HMC are currently under investigation. The algorithm seems to have better update rates, particularly related to the topological tunneling problem.

### 5. Acknowledgements

I would like to thank Brian Pendleton for useful discussions. I am grateful to PPARC for financial support. The numerical work was carried out on DEC Alpha-workstations in Edinburgh (PPARC grant GR/J59142).